\documentclass[12pt]{iopart}
\usepackage{citesort}

\usepackage{inputenc}
\usepackage{graphicx}
\usepackage{subfigure}
\usepackage{color}
\usepackage{cite}
\usepackage{setspace}

\begin{document}

\title[Magnetic merging of ultracold atomic gases of $^{85}$Rb and $^{87}$Rb]{Magnetic merging of ultracold atomic gases of $^{85}$Rb and $^{87}$Rb}

\author{S H\"{a}ndel\footnote[1]{These authors have contributed equally to this work.}, T P Wiles$\dagger$, A L Marchant, S A Hopkins, C S Adams and S L Cornish}

\address{Department of Physics, University of Durham, South Road, Durham DH1 3LE, UK}
\ead{sylvi.haendel@dur.ac.uk}
\begin{abstract}
We report on the magnetic merging of ultracold atomic gases of $^{85}$Rb and $^{87}$Rb by the controlled overlap of two initially spatially-separated magnetic traps. We present a detailed analysis of the combined magnetic field potential as the two traps are brought together which predicts a clear optimum trajectory for the merging. We verify this prediction experimentally using $^{85}$Rb and find that the final atom number in the merged trap is maximised with minimal heating by following the predicted optimum trajectory. Using the magnetic merging approach allows us to create variable ratio isotopic Rb mixtures with a single laser cooling setup by simply storing one isotope in a magnetic trap before jumping the laser frequencies to the transitions necessary to laser cool the second isotope. This offers a simple and cost effective way to achieve suitable starting conditions for sympathetic cooling of $^{85}$Rb by $^{87}$Rb towards quantum degeneracy.
\end{abstract}

\maketitle


\section{Introduction}
The field of matterwave optics has progressed enormously since the first observations of Bose-Einstein condensation~\cite{Anderson1995,Davis1995,Bradley1995} in 1995 and now many more atomic elements are candidates for study in the quantum degenerate regime. The production of mixtures of two or more ultracold atomic gases has opened up an exciting field of rich physics; see recent reviews~\cite{Chin2010,Carr2009,Lahaye2009,Bloch2008}.
Ultracold mixtures may be spinor~\cite{Stenger1998,Cornell1998} in nature, isotopic mixtures~\cite{Papp2006,Zhang2005}, or mixtures of two~\cite{Chin2010,Carr2009} or three~\cite{Jing2008} different atomic species.
Mixtures of the majority of the various isotopic combinations of Li, Na, K, Rb and Cs have been prepared and studied and are conveniently tabulated in Table IV of~\cite{Chin2010}. Such mixtures open up the intriguing possibility of creating ultracold molecules. It is now commonplace to reversibly create weakly bound molecules from ultracold and quantum degenerate atom pairs of bosons~\cite{Herbig2003,Xu2003,Durr2004}, fermions~\cite{Jochim2003,Greiner2003,Zwierlein2003}, and boson/fermion mixtures~\cite{Ospelkaus2006} using magneto-association at a Feshbach resonance. Moreover, several schemes have now produced ultracold molecules in their rovibrational ground state~\cite{Sage2005,Danzl2008,Lang2008,Ni2008,Viteau2008,Deiglmayr2008}. As well as possessing the above intrinsic interest, mixtures play an important technical role in the sympathetic cooling of `difficult' bosonic species such as $^{85}$Rb~\cite{Papp2008,Altin2010} and $^{41}$K~\cite{Thalhammer2008} and all fermions~\cite{Ketterle2008} owing to the suppression of s-wave scattering for fermions.

A mixture of $^{85}$Rb and $^{87}$Rb has several attractive features. It has two interspecies Feshbach resonances~\cite{Burke1998} at $265\,\mbox{G}$ and $372\,\mbox{G}$ which have been used to produce heteronuclear molecules~\cite{Papp2006}. The interspecies elastic cross-section is favourable for sympathetic cooling of $^{85}$Rb \cite{Burke1998}, initially demonstrated in 2001~\cite{Bloch2001} and later used to reach quantum degeneracy by two groups~\cite{Papp2008,Altin2010}. The broad intraspecies Feshbach resonance in $^{85}$Rb has been extensively used to control the atomic interactions in a Bose-Einstein condensate~\cite{Cornish2000}, permitting the study of the collapse of a condensate~\cite{Roberts2001a,Donley2001} and the formation of bright matter-wave solitons~\cite{Cornish2006}, as well as enabling the investigation of phase separation in a dual--species $^{85}$Rb -- $^{87}$Rb condensate~\cite{Papp2008}.

\begin{figure}
	\centering
		\includegraphics[width=0.7\textwidth]{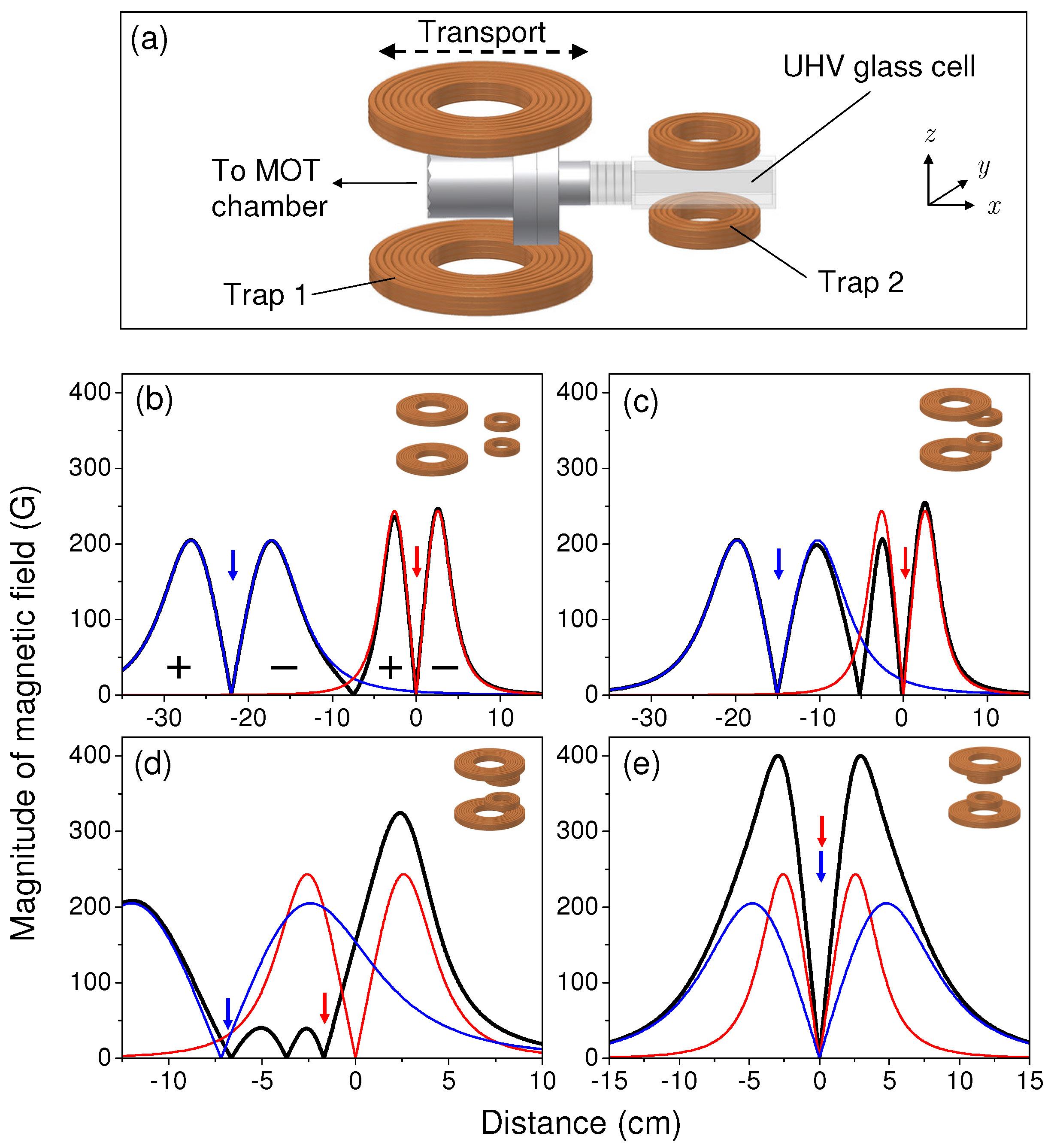}
	\caption{(a) Schematic of the experiment. A cloud of ultracold atoms is transported from the MOT chamber in a quadrupole trap (trap 1) mounted on a motorized translation stage.  Trap 1 reaches the UHV glass cell and the cloud is transferred into the static quadrupole trap (trap 2). Trap 1 then returns to the MOT chamber, collects a second cloud and is merged with the static trap. (b)-(e) The magnitude of the magnetic field along the x-axis for trap 1 (blue), trap 2 (red) and the sum (black) for different separations of the coils. (b) Separation 22.5~cm: Two separate quadrupole traps. The signs (+/-) indicate the direction of the field. (c) Separation 15~cm: As trap 1 approaches trap 2 an additional quadrupole-like zero is created where the red and the blue curves cross and the opposing signs of the field cause cancellation. The two inner barriers in the magnetic potential prevent the atoms entering the central trap. (d) Separation 7.5~cm: The height of the inner barriers is significantly reduced as the separation of the traps is decreased. However, the atoms are still confined in the two outer traps. Note the gradient ratio has been adjusted between (c) and (d) to maintain two inner barriers of the same height. (e) Separation 0~cm: Once merging is complete, both traps are overlapped to create a single quadrupole trap.}
	\label{fig:CoilSchematic2}
\end{figure}

One problem encountered when preparing a mixture of two species is the need to duplicate two sets of lasers (cooling and repumping) and optics for the initial cooling and trapping stages. Apart from the expense, the footprint of all the necessary devices is considerable and the proliferation of components can reduce the available optical access for the final stages of the experiment. In this paper we report an alternative scheme to prepare isotopic mixtures using a single laser cooling setup and two magnetic traps which are controllably merged to combine the two atomic gases. A magnetic merging scheme has been demonstrated once before~\cite{Jesper2007}, but only for a single atomic species. Here we demonstrate that the approach may be used to create ultracold atomic mixtures and provide detailed insight into the merging process. The overall sequence of our experiment is as follows. Ultracold atoms are collected from a background vapour in a magneto-optical trap (MOT), loaded into a magnetic quadrupole trap (trap 1) and transported~\cite{Lewandowski2003} from the MOT chamber to a UHV glass cell (figure~\ref{fig:CoilSchematic2}(a)). The atoms are then transferred into a static quadrupole trap (trap 2) and trap 1 is translated back to the MOT chamber. At the same time, if desired, the laser frequencies can be jumped to the transitions necessary to laser cool the second isotope. A second sample of atoms is collected and again transported to the UHV cell where the two traps are controllably merged. The merging of the two traps is non-trivial and the bulk of this paper is devoted to a detailed theoretical and experimental study of the merging process. In particular, we provide a detailed analysis of the combined magnetic field potential during the merging process which highlights the optimum merging trajectory. The analysis should have general applicability for any similar trap-merging experiments.

The structure of the paper is as follows. In section~\ref{sec:Theory} we describe the calculations of the magnetic potential which lead to the prediction of an optimum merging trajectory. We present our experimental apparatus in section~\ref{sec:ExperimentalSetup}. In section~\ref{sec:ExperimentalOptimisation} we describe experiments that test the predictions of our magnetic potential analysis, and verify that the predicted optimum trajectory maximises the final atom number in the merged trap with minimal heating of the gas. Finally we demonstrate merging of variable proportions of the two different isotopes of rubidium.


\section{Theory}\label{sec:Theory}

In this section we analyse the combined magnetic potential as the two quadrupole traps are brought together, showing that there is an optimum trajectory for successful merging.

\subsection{Modelling the quadrupole traps}\label{sec:model}

The magnetic field generated by a coil is calculated using the Biot-Savart law

\begin{equation}
\centering
\mathbf{B}=\int{\frac{\mu_{0}I'}{4\pi}\frac{\rm{d}\mathbf{l}\times\mathbf{r}}{\left|\mathbf{r}\right|^{3}}},
\label{eq:biotsavart}
\end{equation}
where $\mu_{0}$ is the permeability of free space, $I'$ is the current through the coil, $\rm{d}\mathbf{l}$ is an infinitesimally small element of the coil and $\mathbf{r}$ is the vector from the element $\rm{d}\mathbf{l}$ to the point in space where the magnetic field is to be calculated. In the experiment the real coils are wound from multiple turns of square cross-section copper tubing with the dimensions summarised in table~\ref{tab:tabone}. For simplicity, the coils are approximated by `equivalent coils' consisting of a single turn of infinitesimal thickness and carrying a current of $I'= N\cdot I$, where $N$ is the number of turns of the real coil and $I$ is the current in the real coil.  The radii and separations of the `equivalent coils' were found by matching the calculated first and third spatial derivatives to the measured values for the real coils. Comparing the measured and calculated magnetic fields results in a normalized RMS deviation of $\approx$~1\% over the range of interest, confirming the validity of this approximation.

\begin{table}
\caption{\label{tab:tabone}Measured dimensions of the coils and their `equivalent coil' parameters. All the coils are wound from square cross-section copper tubing.}
\begin{indented}
\lineup
\item[]\begin{tabular}{@{}*{6}{l}}
\br
 & Trap 1 & Trap 2\cr
\mr
			Number of turns & 3 $\times$ 8 & 3 $\times$ 3\cr
			Tubing dimensions~(mm $\times$ mm) & 4.0 $\times$ 4.0 & 3.5 $\times$ 3.5 \cr
			Inner separation~(cm) & 8.6(1) & 3.7(1)\cr
			Outer separation~(cm) & 11.1(1) & 5.9(1)\cr
			Inner radius~(cm) & 3.0(1) & 2.2(1)\cr
			Outer radius~(cm) & 6.5(1) & 3.4(1)\cr\hline
			Equivalent coil separation~(cm) & 10.4(1) & 4.7(1)\cr
			Equivalent coil radius~(cm) & 4.9(1) & 2.7(1)\cr
			Axial field gradient~($\,\mbox{G$\,$cm$^{-1}\,$A$^{-1}$}$) & 0.606(1) & 0.974(1)\cr
			Field maximum~(G$\,$A$^{-1}$) & 1.004 & 0.961\cr
			\br
\end{tabular}
\end{indented}
\end{table}

\subsection{Modelling the merging process}\label{sec:process}

In order to understand the merging process the combined magnetic field resulting from both traps is calculated as a function of their separation. Due to adiabatic following, the magnetic potential, $U_{\rm{mag}}$, experienced by an atom is proportional to the magnitude of the field, \textit{i.e.} $U_{\rm{mag}}=m_{\rm{F}}g_{\rm{F}}\mu_{\rm{B}}\left|B\right|$ \cite{Migdall1985}, where $m_{\rm{F}}$ is the magnetic sub-level, $g_{\rm{F}}$ is the Land\'{e} g-factor and $\mu_{\rm{B}}$ is the Bohr magneton. We therefore calculated the combined magnetic potential on a three-dimensional grid. From this grid we were able to generate one-dimensional cuts, two-dimensional contours and three-dimensional isosurfaces of this potential. A preliminary analysis revealed that the essential details of the merging process could be extracted from the simpler one-dimensional cuts of the combined magnetic potential along the line joining the two trap centres.

Examples of such one-dimensional cuts are shown in figures~\ref{fig:CoilSchematic2}(b-e), where the magnetic field due to traps 1 and 2 and the combined magnetic field are indicated by the blue, red and black lines respectively. For a given set of coils the form of the combined magnetic potential depends critically on two parameters; the separation of the two trap centres and the ratio of the axial magnetic field gradients at each trap centre (trap 2/trap 1), henceforth referred to as the `gradient ratio'. In figures~\ref{fig:CoilSchematic2}(b-e) we follow the condition that the inner barriers of the magnetic potential that separate the two traps are maintained at equal heights. This requires the gradient ratio to be adjusted as the trap centres approach each other.

\subsection{Theoretical results}\label{sec:theoreticalresults}

\begin{figure}
	\centering
		\includegraphics[width=0.75\textwidth]{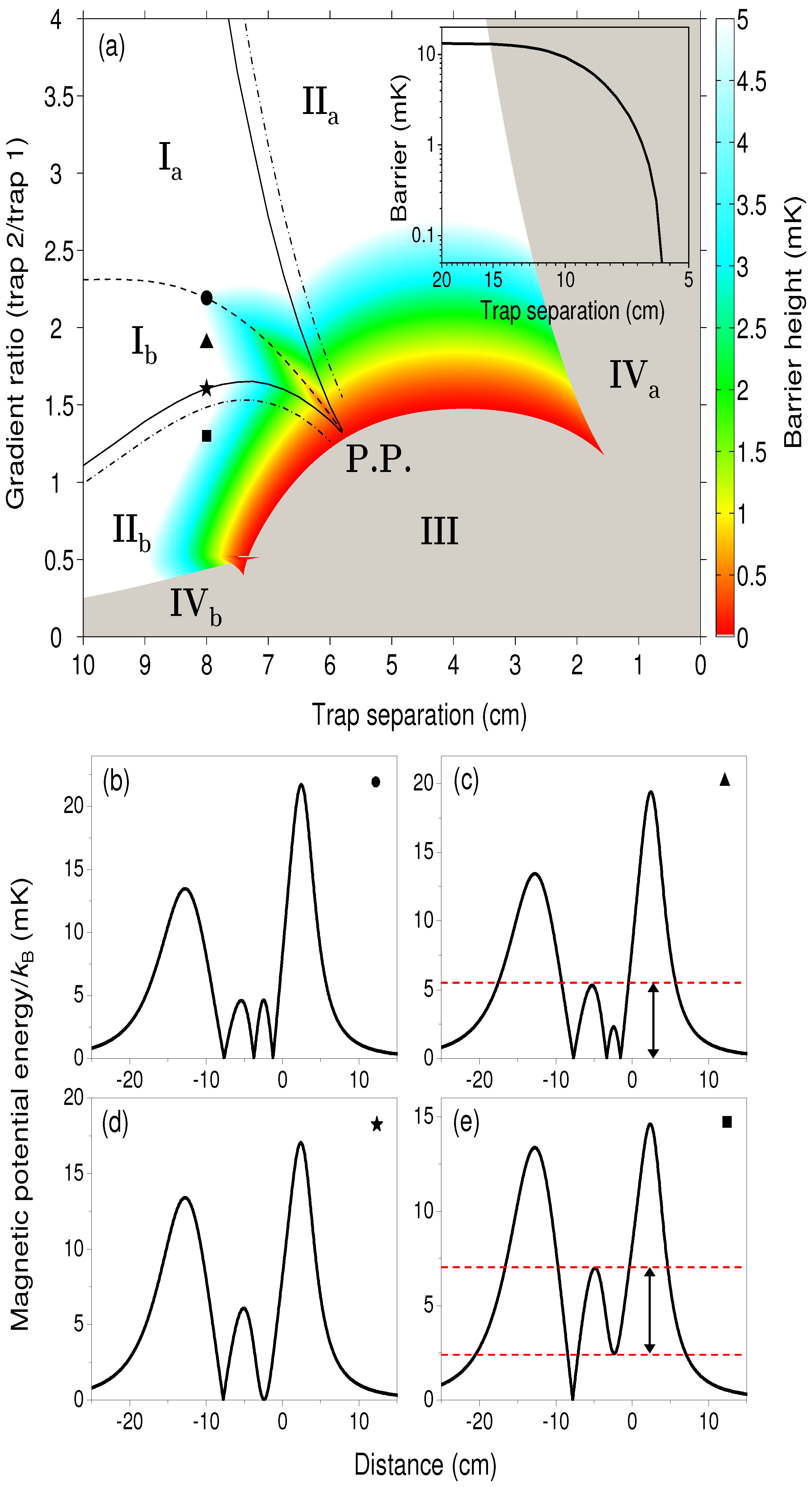}
	\caption{(a) Combined magnetic potential as a function of gradient ratio (trap 2/trap 1) and trap separation for $^{85}$Rb ($F = 2, m_{\rm{F}} = - 2$), where the axial field gradient of trap 1 is $180\,\mbox{G$\,$cm$^{-1}$}$. The black dashed line indicates where the two inner potential barriers are maintained at an equal height and the inset of (a) shows how this barrier height changes as a function of the trap separation. The black solid lines show the trajectories where two zeros have merged and the resulting single minimum is about to lift up. The upper (lower) dot-dashed line represents where the trap 1 (trap 2) minimum is lifted to 1 mK. (b-e) show the combined potentials along the transport axis for the points marked in (a). Red dashed lines in (c) and (e) indicate the relevant barrier heights plotted in (a) as a colour variation in mK for regions \textbf{I} and \textbf{II}, respectively. The `pinch point' referred to in the main text is labelled as `\textbf{P.P.}'.}
	\label{fig:colormap}
\end{figure}

To fully explore the merging process a set of $\approx$~1000 one-dimensional plots of combined magnetic potential was generated for varying trap separations and gradient ratios. To condense this grid of plots into a more useful form we identified regions where qualitatively similar behaviour was present in the combined magnetic potential, such as the presence of three field zeros or the existence of a single merged trap. These regions are shown in figure~\ref{fig:colormap}(a) as a function of trap centre separation and gradient ratio. Each merging event has a unique trajectory through this `potential map', travelling from left to right on the figure. Examples of four individual points in figure~\ref{fig:colormap}(a) are shown in figures~\ref{fig:colormap}(b-e) in order to demonstrate how the combined magnetic potential changes with respect to gradient ratio for a fixed trap separation.


Understanding the potential map is fundamental to understanding the merging process, therefore, we will now discuss in detail the regions highlighted in figure~\ref{fig:colormap}(a). Regions with differing Roman numerals identify different qualitative behaviour in the combined magnetic potential. In region \textbf{I} the combined magnetic potential exhibits three field zeros separated by two potential barriers (figures~\ref{fig:colormap}(b) and (c)). Separating this region into two parts is the dashed line, along which the barriers are of equal height (figure~\ref{fig:colormap}(b)). The barrier height along this line is shown, as a function of separation, in the inset of figure~\ref{fig:colormap}(a). Either side of the dashed line the barriers are asymmetric and smaller on the side of trap 1 (\textbf{I$_{\rm{a}}$}) or trap 2 (\textbf{I$_{\rm{b}}$}) (figure~\ref{fig:colormap}(c)). In region \textbf{II} there are only two field minima and one barrier as the central field zero and either trap 1 (\textbf{II$_{\rm{a}}$}) or trap 2 (\textbf{II$_{\rm{b}}$}) have combined (figures~\ref{fig:colormap}(d) and (e)). The solid black lines mark the boundary between regions \textbf{I} and \textbf{II} (figure~\ref{fig:colormap}(d)). As we venture further into region \textbf{II} the magnetic field of the combined minimum becomes non-zero and increases as we move further from the solid line (figure~\ref{fig:colormap}(e)). The dot-dashed lines indicate where this minimum has been lifted to a potential of 1~mK. Note that all potential energies presented in this analysis refer to $^{85}$Rb ($F = 2, m_{\rm{F}} = - 2$) and an axial field gradient of trap 1 equal to $180\,\mbox{G$\,$cm$^{-1}$}$. 

The optimum merging strategy is to avoid the raised minima that occur in region \textbf{II} as the raised potential could heat the cloud. Consequently the optimal `merging channel' is given by the boundaries of region \textbf{I}, where two intermediate barriers and three magnetic field zeros continue to be maintained. The three lines guiding the channel converge and lead into region \textbf{III} at the `pinch point' (indicated in figure~\ref{fig:colormap}(a) as \textbf{P.P.}). In region \textbf{III} (and region \textbf{IV}) traps 1 and 2 have merged into a single trap. In this simple picture we identify optimal trajectories as those which broadly follow the dashed line in region \textbf{I} passing through the merging channel and entering region \textbf{III} at the pinch point, thereby avoiding the raised minima in region \textbf{II}.


The colour map in figure~\ref{fig:colormap}(a) represents the relevant barrier height and can be used to assess the extent of merging between the two atom clouds with finite energy. While the merging process is not adiabatic, to give some indication of the kinetic energy of atoms in the traps with respect to the inner barrier heights, we will treat the clouds as being in thermal equilibrium in the following analysis. We assume a temperature of $250\,\mbox{$\mu$K}$ for a gas confined in a trap with an axial field gradient of $180\,\mbox{G$\,$cm$^{-1}$}$. In region \textbf{I} the colour map indicates the maximum intermediate barrier height in mK, given by the red dashed line in figure~\ref{fig:colormap}(c). Within the white area of this region no merging occurs as $<1\,\mbox{\%}$ of atoms have sufficient kinetic energy to traverse potential barriers of height $>2.75\,\mbox{mK}$. Merging of the two traps begins to occur when the trajectory enters the green area, however, it is important to note that atoms from one trap may spill over the smaller barrier into the intermediate trap while the combined magnetic potential remains in region \textbf{I}. Once the trajectory reaches the yellow area $\approx50\,\mbox{\%}$ of atoms can traverse the inner barriers. In region \textbf{II} the colour map shows the minimum relative barrier height, given by the red dashed lines in figure~\ref{fig:colormap}(e). Here atoms could suffer an undesirable gain in kinetic energy during merging as they may be dropped from the lifted non-zero minimum into the trap with a field zero. Introducing finite energy broadens the merging channel because individual traps can be lifted if the kinetic energy that results from dropping the atoms into the second trap is small in comparison with the initial thermal energy of the atoms. The dot-dashed lines in figure~\ref{fig:colormap}(a) give an indication of this broadening for a tolerance of the trap lifting up to $1\,\mbox{mK}$.

In order to confirm the validity of the above analysis we identify two trajectories on figure~\ref{fig:colormap}(a) that we test experimentally in section~\ref{sec:ExperimentalOptimisation}. The first of these trajectories is maintaining a constant gradient ratio during merging (see figure~\ref{fig:PlainQT}). We predict that for high or low gradient ratios atoms will only be maintained from the dominant trap and expect that poor merging will occur for intermediate ratios due to these not following the optimal trajectory. For our second trajectory we aim to maintain a gradient ratio that follows the dashed line on figure~\ref{fig:colormap}(a). Since merging only begins to occur in the green area of region \textbf{I}, we choose a trajectory consisting of a straight line followed by a ramp downward tangential to the dashed line within the red, yellow and green areas (see figure~\ref{fig:OverviewRamps}(a) inset). We expect to observe optimal merging following such trajectories.


\section{Experimental setup}\label{sec:ExperimentalSetup}
In this section we present a brief description of the experimental setup. The apparatus is divided between two independent optical tables. The first table houses the laser system, which is used to prepare the necessary light frequencies for laser cooling, repumping, optical pumping and imaging. The light is delivered by optical fibres to the second table on which the vacuum system, laser cooling and magnetic trapping hardware is situated. A key feature of the apparatus is that, at any one time, the laser system only generates the light for \emph{either} $^{85}$Rb or $^{87}$Rb, but the system can be easily switched from one isotope to the other during the course of an experimental run.

The laser setup consists of two commercial extended cavity diode lasers (Toptica DL100) and a tapered amplifier (Toptica BoosTA). Both the diode lasers operate on the 780\,nm 5S$_{1/2}$ $\rightarrow$ 5P$_{3/2}$ transition. The first laser generates the light for laser cooling and imaging and is stabilized to the `cycling' transition ($F = 3 \rightarrow$ $F' = 4$ for $^{85}$Rb or $F = 2 \rightarrow$ $F' = 3$ for $^{87}$Rb) using modulation transfer spectroscopy~\cite{MCCarron2008}. The necessary variable detunings required for laser cooling and absorption imaging are generated using several acousto-optical modulators (AOMs) in a double-pass configuration. The AOMs also allow real-time control of the light intensity. The tapered amplifier is used to increase the amount of light available for laser cooling. The second laser generates the light for repumping and optical pumping and is stabilized to the `repump' transition ($F = 2 \rightarrow F' = 3$ for $^{85}$Rb or $F = 1 \rightarrow F' = 2$ for $^{87}$Rb) using frequency-modulation spectroscopy~\cite{Gehrtz1985}. Again AOMs are used for intensity control of the repump light. An additional AOM is used in a double (single) pass configuration to generate the optical pumping light for $^{85}$Rb ($^{87}$Rb).

This versatile setup allows us to produce ultracold atomic gases of $^{85}$Rb or $^{87}$Rb by simply relocking the two extended cavity diode lasers to the equivalent transitions. This simple switch is only possible because the lasers are locked directly to the `cycling' and `repump' transitions. Alternative schemes, for example stabilizing the laser frequency to a cross-over resonance, require fewer AOMs but cannot be simply switched from one isotope to the other.

The vacuum system is divided into two sections connected by a differential pumping stage. Ultracold atoms are prepared in a standard six-beam MOT configuration~\cite{Raab1987} in a stainless steel octagonal chamber. Here we typically load up to $10^9\,^{85}$Rb atoms (or $7\times10^8\,^{87}$Rb atoms) in less than $10\,\mbox{s}$ from a background vapour of rubidium supplied by a dispenser (SAES Getters). After a short compressed MOT phase and a molasses stage~\cite{Chu1985}, we optically pump the atoms into the $F = 2, m_{\rm{F}} = - 2$ state for $^{85}$Rb or the $F = 1, m_{\rm{F}} = - 1$ state for $^{87}$Rb. The atoms are then loaded into a quadrupole trap (trap 1) with an axial field gradient of $50\,\mbox{G$\,$cm$^{-1}$}$. The gradient is then adiabatically increased to $160\,\mbox{G$\,$cm$^{-1}$}$ before the trap is transported over $50\,\mbox{cm}$ along the vacuum system in $2.5\,\mbox{s}$ to an UHV glass cell. The magnetic transport~\cite{Lewandowski2003} is achieved by mounting the quadrupole trap on a motorized translation stage (Parker 404 series) which has a positioning accuracy of $5\,\mbox{$\mu$m}$. Movement of the translation stage can be programmed to follow a variety of velocity profiles, with accurate control of the speed and acceleration. Further details of the magnetic transport and the construction of the apparatus will be presented elsewhere.

Once in the UHV cell, the atoms are transferred to a second static quadrupole trap (trap 2). This is achieved by first fully overlapping the two sets of quadrupole coils and then adiabatically turning on the gradient of trap 2 to $320\,\mbox{G$\,$cm$^{-1}$}$ whilst simultaneously decreasing the gradient of trap 1 to zero. The lifetime of the trapped gas in trap 2 is $(240\pm10)\,\mbox{s}$ and the observed heating rate is $0.30(2)\,\mbox{$\mu$K$\,$s$^{-1}$}$. Having transferred the atoms to trap 2, trap 1 is returned to the MOT chamber to collect a second cloud of atoms, which is again transported to the UHV cell. The two trapped samples are then controllably merged by overlapping the two sets of coils, this time with currents flowing in both sets of coils. When the centre of trap 1 is $10\,\mbox{cm}$ away from the centre of trap 2 the hardware controlling the motorized translation stage generates a trigger which is read by the main experimental control system. This allows subsequent ramps of the magnetic field gradients to be precisely timed with respect to the motion of trap 1. Following such a merging sequence, the currents generating both traps are switched off in $<0.2$\,ms and standard absorption imaging techniques are used to probe the temperature and density of the combined atomic cloud.

When merging experiments with two different isotopes are performed, we initially collect $^{85}$Rb and then switch the laser frequencies as trap 1 returns to the MOT chamber in order to collect an equal number of $^{87}$Rb atoms. To compensate partially for the difference of magnetic moment between $^{85}$Rb $(m_{\rm{F}}g_{\rm{F}} = 2/3)$ and $^{87}$Rb $(m_{\rm{F}}g_{\rm{F}} = 1/2)$, we increase the gradient of trap 1 whilst transporting $^{87}$Rb to $205\,\mbox{G$\,$cm$^{-1}$}$ (limited by heating in the coils).

\section{Experimental results}\label{sec:ExperimentalOptimisation}
To test the predictions based upon the potential map in figure~\ref{fig:colormap}(a), we initially performed a series of merging experiments with $^{85}$Rb atoms confined in both traps (sections \ref{sec:fixedgrad}-\ref{sec:Optimisedmerging}). Subsequently we demonstrate the merging of two different isotopes of rubidium (section \ref{sec:merge8587}).

\subsection{Merging with fixed field gradients}\label{sec:fixedgrad}
\begin{figure}
	\centering
		\includegraphics[width=0.88\textwidth]{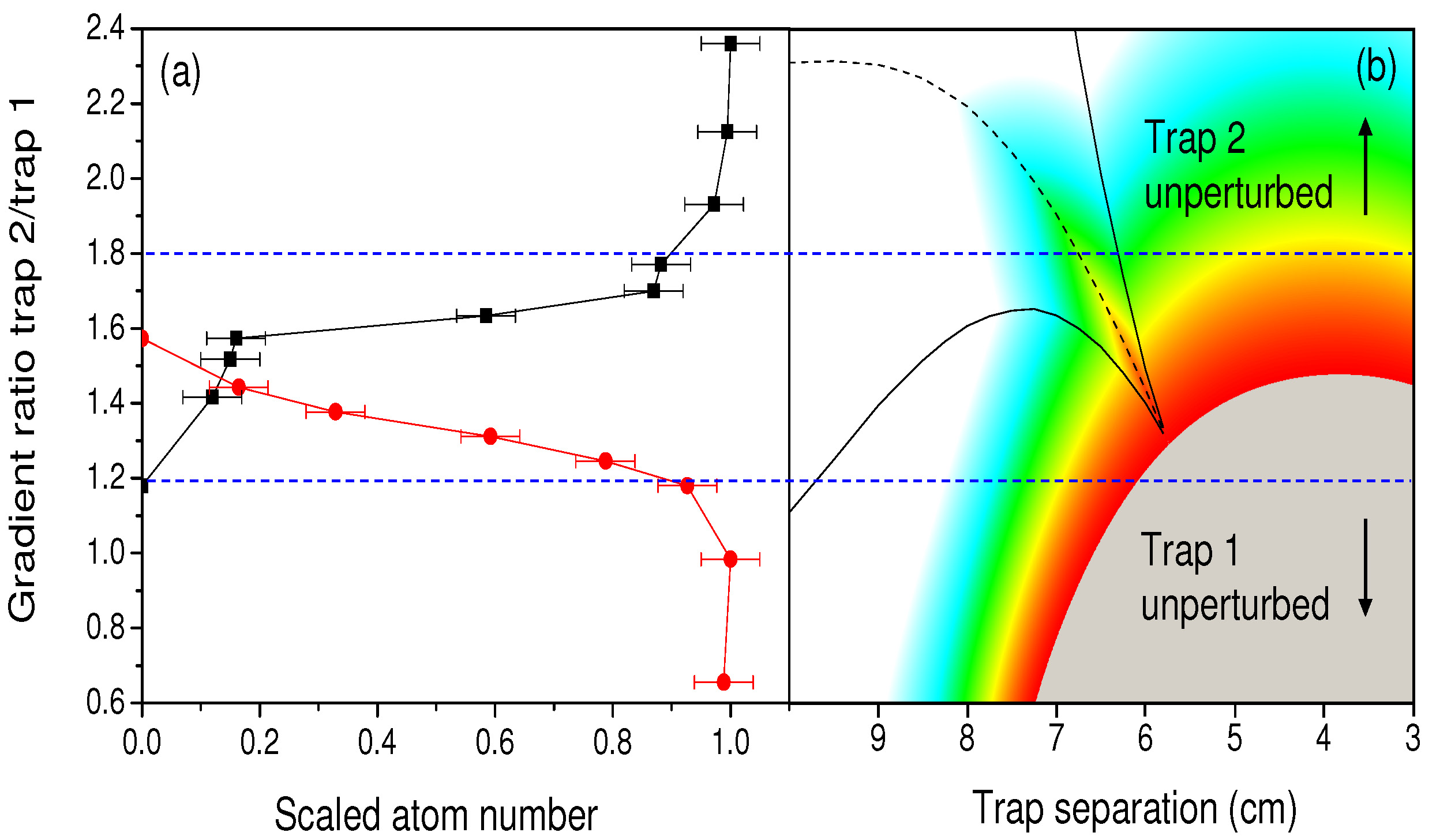}
			\caption{(a) Scaled atom number after merging as a function of the fixed gradient ratio of the two traps. The atoms are initially confined either in trap 2 (black squares) or in trap 1 (red circles). In both cases the second trap is initially empty. The blue dashed lines indicate the gradient ratios for each trap where $10\,\mbox{\%}$ of the atoms are lost during a merge event. These lines are also shown in the potential map in (b) to highlight the regions where either trap 1 or trap 2 is largely unperturbed by the presence of the other trap.}
	\label{fig:PlainQT}
\end{figure}

Our initial aim was to investigate the first trajectory identified in section~\ref{sec:theoreticalresults}; that is whether merging could be achieved for a fixed value of the gradient ratio. To test this we confined the atoms initially in either trap 1 or trap 2 and then merged the two traps with a velocity of $5\,\mbox{cm$\,$s$^{-1}$}$ for a range of constant gradient ratios. In both cases the second trap was initially empty. The results of these simple experiments are shown in figure~\ref{fig:PlainQT}. Throughout the paper the measured atom number is scaled to the maximum number loaded into each trap before merging. For the results shown in figure~\ref{fig:PlainQT}, this corresponded to $(5.3\pm0.3)\times 10^8$ for trap 1 and $(7.0\pm0.4)\times 10^8$ for trap 2. The regions above and below the blue dashed lines identify the gradient ratios where $<10\,\mbox{\%}$ of the atoms are lost from either trap during the merging process and these lines are also indicated on the potential map. Henceforth, we shall refer to the dominant trap in each of these regions as being `unperturbed'. As predicted in section~\ref{sec:Theory}, at high gradient ratios trap 2 dominates trap 1. Evidence of this can be seen in figure~\ref{fig:PlainQT}(a) where atoms from the weaker trap 1 fail to enter the stronger trap 2. The converse can be seen for low gradient ratios. Also, as predicted, there is a smooth transition between the two unperturbed regions. However, the merging here is highly inefficient. For example, at a gradient ratio of $\approx1.5$, where the atom number is equal for both traps, only $\approx15\,\mbox{\%}$ of the atoms are retained from each of the traps.

\subsection{Merging with ramped field gradients}
\begin{figure}
	\centering
		\includegraphics[width=0.57\columnwidth]{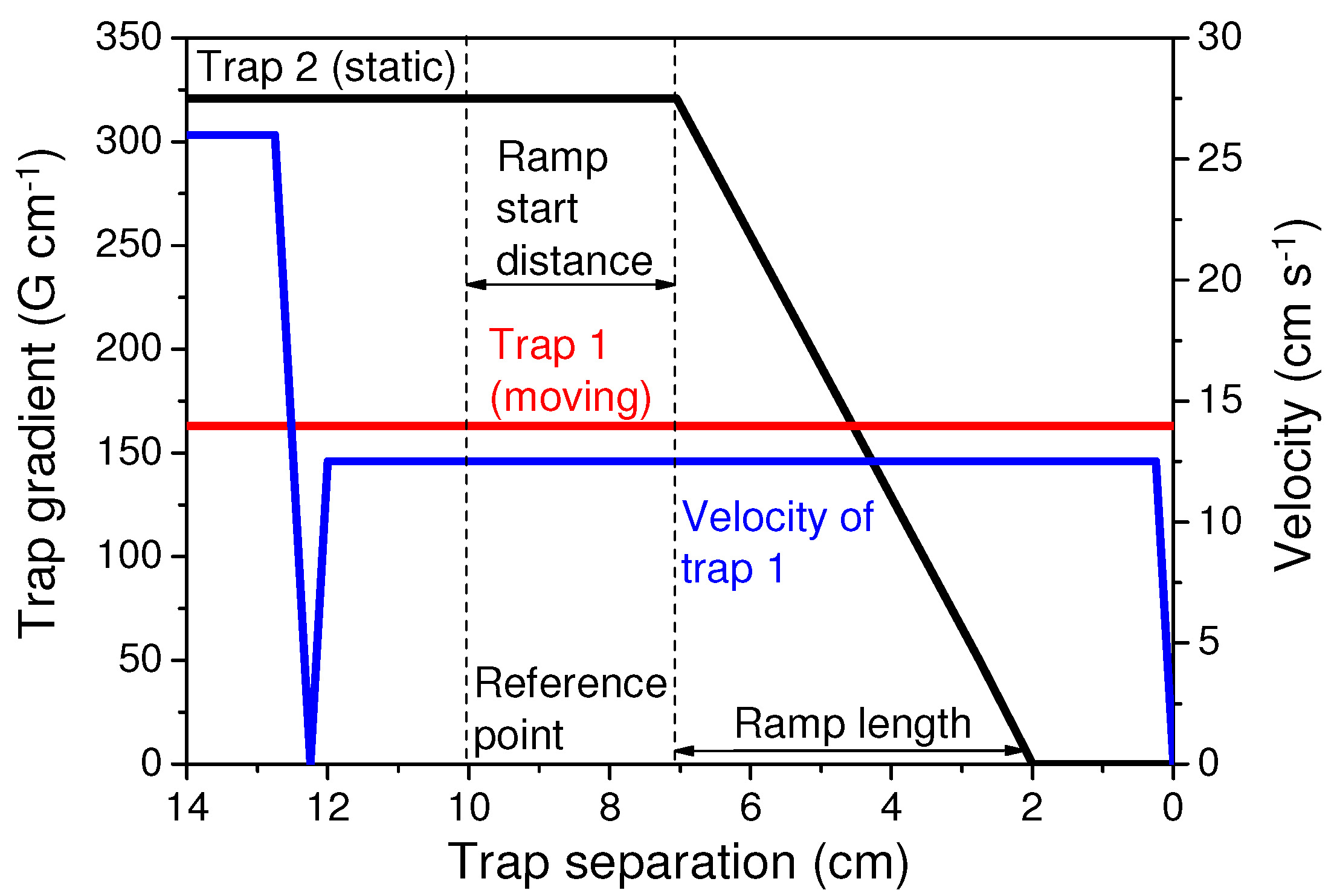}
		\caption{Typical evolution profiles of the gradient and velocity of the quadrupole traps during merging. The gradient of trap 1 (red) remains constant, whilst the gradient of trap 2 (black) is decreased during the merging process. The blue line shows the velocity profile of trap 1, which decreases from the transport setting to a new variable velocity before the merging begins.}
	\label{fig:CurrentRampSketch}
\end{figure}

In order to discover whether successful merging could be achieved by following the narrow merging channel of region \textbf{I}, we employed a linear ramp of the gradient ratio during the merging as depicted in figure~\ref{fig:CurrentRampSketch}. In this scenario the gradient in trap 1 is held constant at $160\,\mbox{G$\,$cm$^{-1}$}$ as it is moved towards the static trap 2. At a separation of $10\,\mbox{cm}$, where the traps are far from the merging region (see figure~\ref{fig:colormap}(a)), the reference trigger from the hardware controlling the motorized translation stage is sent to our control system. After trap 1 has travelled a variable distance beyond this reference point, which we refer to as the `ramp start distance', the gradient of trap 2 is linearly decreased from the initial $320\,\mbox{G$\,$cm$^{-1}$}$ to zero over a variable time. In order for the data to be comparable to figure~\ref{fig:colormap}(a) we relate the time taken for this linear ramp to the distance that the trap separation has decreased in this time and define it as `ramp length'. The velocity of trap 1 is altered from the transport setting ($26\,\mbox{cm$\,$s$^{-1}$}$) to a new variable velocity before the merging begins, as depicted in figure~\ref{fig:CurrentRampSketch}. By varying the ramp start distance and the ramp length we are able to explore the potential map in figure~\ref{fig:colormap}(a), searching for the optimum merging trajectory.

\begin{figure}
	\centering
		\includegraphics[width=1\columnwidth]{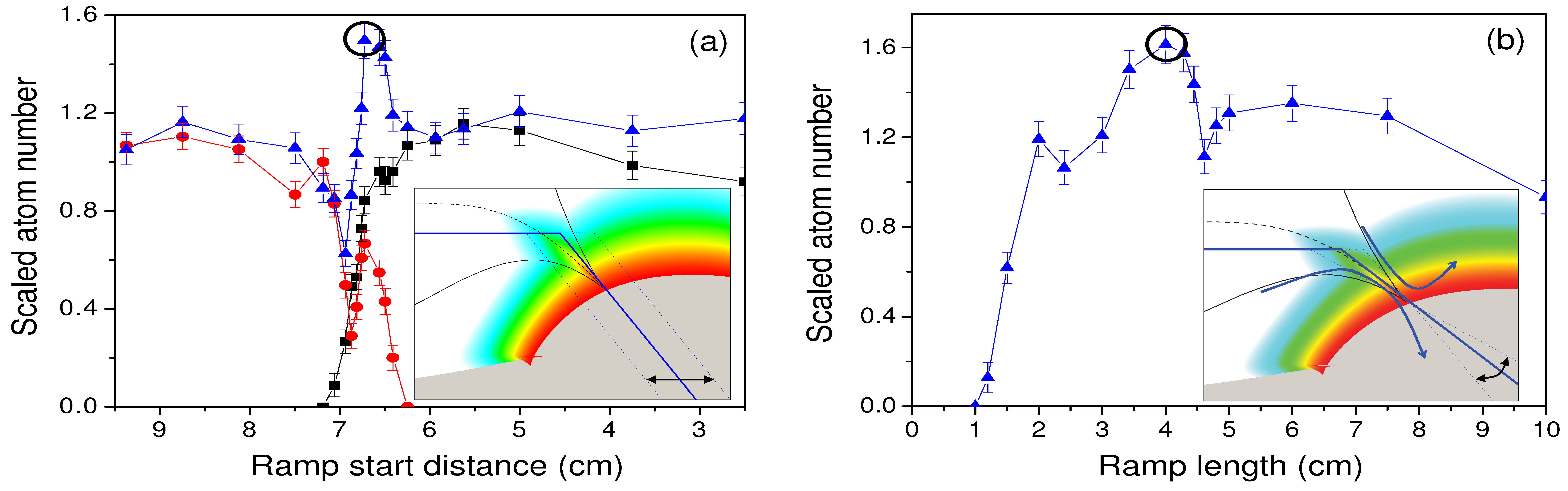}
	\caption{Optimising the merging process through measurements of the scaled atom number following the merging as a function of the ramp parameters. In (a) we vary the ramp start position, for a ramp length of $3.0\,\mbox{cm}$, translating the ramp across the pinch point and in (b) we vary the ramp length \textit{i.e.} the angle of the trajectory through the pinch point. Data are shown for atoms initially in trap 2 (black squares), trap 1 (red circles) and both traps (blue triangles). The solid blue lines in the insets indicate the ramps circled in each figure, while the dotted blue lines indicate the variation of the ramp in each experiment. The results of a large number of experimentally tested ramps are summarised in the inset of (b) with the curved blue arrows indicating where the merged atom number is $\approx50\,\mbox{\%}$ of that seen for the optimal trajectory.}
	\label{fig:OverviewRamps}
\end{figure}

In a first set of experiments, we fixed the ramp length and varied the ramp start distance, thus translating the ramp horizontally (w.r.t to figure~\ref{fig:colormap}(a) and figure~\ref{fig:CurrentRampSketch}) across the merging channel and the pinch point. For each experiment the merging was performed three times; firstly with the atoms in trap 1, then with the atoms in trap 2 and finally with the atoms loaded into both traps. The results in figure~\ref{fig:OverviewRamps}(a) are for a ramp length of $3\,\mbox{cm}$, merging speed of $12.5\,\mbox{cm$\,$s$^{-1}$}$ and inital atom number of $1.5\times10^{8}$. The left hand side of this figure corresponds to an experiment where effectively trap 2 is turned off before trap 1 arrives, and accordingly all the atoms initially in trap 1 remain. The right hand side of the figure corresponds to an experiment where trap 1 and trap 2 are effectively merged with a constant gradient ratio (equal to 2 in this case). In this limit, therefore, the results are consistent with the experiment presented in figure~\ref{fig:PlainQT}, with the majority of the atoms initially in trap 2 remaining. In the central region of the figure a mixture of atoms from trap 1 and trap 2 remain in the merged trap. The solid blue line in the inset of figure~\ref{fig:OverviewRamps}(a) depicts the ramp given by the circled data point, which clearly follows the identified merging channel, passing very close to the pinch point. In this case, highly improved merging is observed for an optimum ramp start distance of $6.7\pm0.1\,\mbox{cm}$ when compared to the fixed gradient case shown in figure~\ref{fig:PlainQT}.

We then repeated this experiment for several ramp lengths, with the aim of confirming the existence of the pinch point. In each case we were able to identify an optimum ramp start distance for the given ramp length by requiring that approximately equal numbers of atoms were transferred from each trap into the final merged trap. Strikingly, within experimental error, the optimum ramps intersect and we identify an experimental pinch point of $\mbox{(trap~separation,~gradient~ratio)}=\mbox{($5.8\pm0.1\,\mbox{cm}$,~1.4~$\pm$~0.1)}$. This is in remarkably good agreement with the theoretically determined pinch point of ($5.76\,\mbox{cm}$,~1.30). Moreover, closer examination of thia data highlights that the most successful merging occurs for trajectories that follow the merging channel. These results therefore confirm our predictions that the optimum merging trajectory will follow the merging channel and will pass through the narrow pinch point.

\subsection{Optimising the merging trajectory}\label{sec:Optimisedmerging}
To test the dependence of the merging on the slope of the trajectory, we performed a second set of experiments in which the ramp length was varied whilst the ramp start distance was adjusted to ensure that the trajectory still passed through the experimentally determined pinch point, effectively swivelling around this point (see figure~\ref{fig:OverviewRamps}(b) inset). Traps 1 and 2 were again loaded with equal numbers of atoms and the combined atom number after merging was measured. The results are plotted in figure~\ref{fig:OverviewRamps}(b), identifying an optimum ramp length of $(4.0\pm0.5)\,\mbox{cm}$ through the experimentally determined pinch point. This merging trajectory through the potential map is entirely consistent with our theoretical predictions in section~\ref{sec:theoreticalresults} and there is excellent agreement between the slope of the trajectory and the gradient of the dashed line in figure~\ref{fig:colormap}(a) in the vicinity of the pinch point (see solid blue line in inset of figure~\ref{fig:OverviewRamps}(b)). Note, however, the uncertainty in the optimum ramp length, due to the broad peak seen in figure~\ref{fig:OverviewRamps}(b), is an indication of the presence of a broadened merging channel.

Having established the optimum merging trajectory for the experimentally determined pinch point, we carried out many more runs varying the trajectory around this optimum in order to estimate the size and width of the merging channel. The results are summarized in the inset of figure~\ref{fig:OverviewRamps}(b). The blue curved arrows indicate trajectories where the merged atom number is half of the number obtained along the optimum trajectory. As theoretically predicted in section~\ref{sec:theoreticalresults}, the channel narrows into the crucial pinch point and then opens wide once merging is achieved, with the blue arrows confirming the broadening of the pinch point due to the finite temperature of the atoms.

In a further experiment, we also varied the velocity of trap 1 during the merging for this optimum ramp length. This showed that, within the experimental uncertainties, the maximum achievable atom number is independent of the speed with which the two traps merge up to $12.5\,\mbox{cm$\,$s$^{-1}$}$. For higher speeds the number of atoms remaining in trap 1 falls off quickly towards zero.

\begin{figure}
	\centering
		\includegraphics[width=\columnwidth]{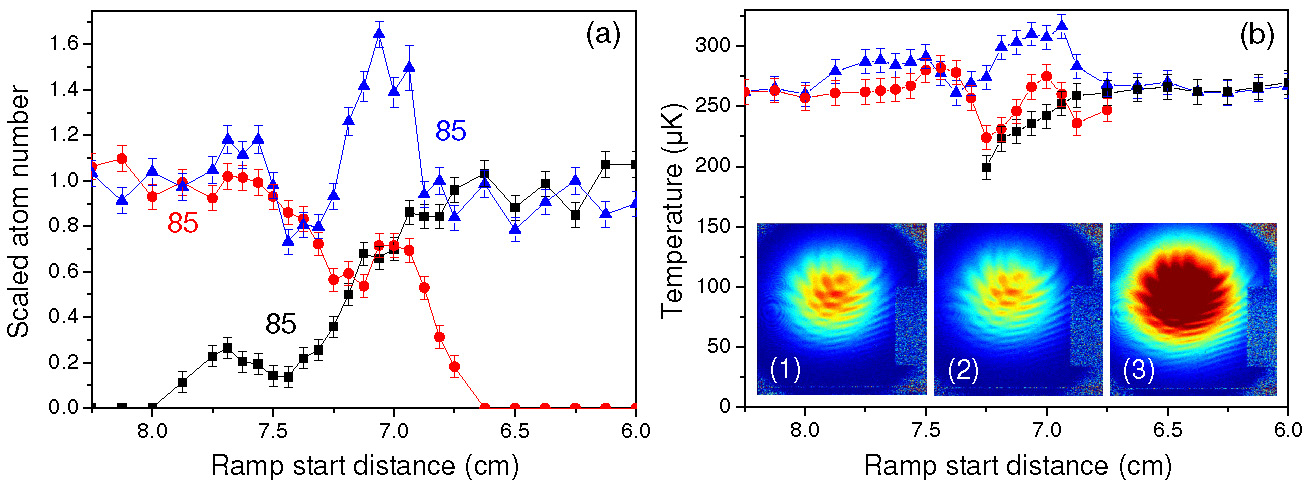}
	\caption{Detailed results of the merging for $^{85}$Rb. Scaled atom number (a) and temperature (b) following the merging (for atoms initially in trap 2 (black squares), trap 1 (red circles) and both traps (blue triangles)) as a function of the ramp start distance for a fixed ramp length of $4.0\,\mbox{cm}$, initial gradient ratio of 2 and a merging speed of 12.5\,cm\,s$^{-1}$. The inset in (b) shows false colour absorption images of the atomic cloud following merging (for atoms initially in (1) trap 1, (2) trap 2 and (3) both traps).}
	\label{fig:AtomnumbervsDistance}
\end{figure}

Using the optimum ramp length determined from figure~\ref{fig:OverviewRamps}(b), we made a detailed measurement of the atom number and the temperature for the merged cloud as a function of the ramp start distance for $^{85}$Rb in both traps (see figure~\ref{fig:AtomnumbervsDistance}). The results indicate that a merge of $\approx75\,\mbox{\%}$ of the atoms from each trap was possible giving $\approx150\,\mbox{\%}$ of the number achieved in a single load. In addition, we do not observe any significant heating with the final merged clouds having typical temperatures of $\approx300\,\mbox{$\mu$K}$ to be compared with the temperatures of the initial clouds of $\approx260\,\mbox{$\mu$K}$.

\subsection{Merging $^{85}$Rb and $^{87}$Rb}\label{sec:merge8587}

\begin{figure}
	\centering
		\includegraphics[width=0.5\columnwidth]{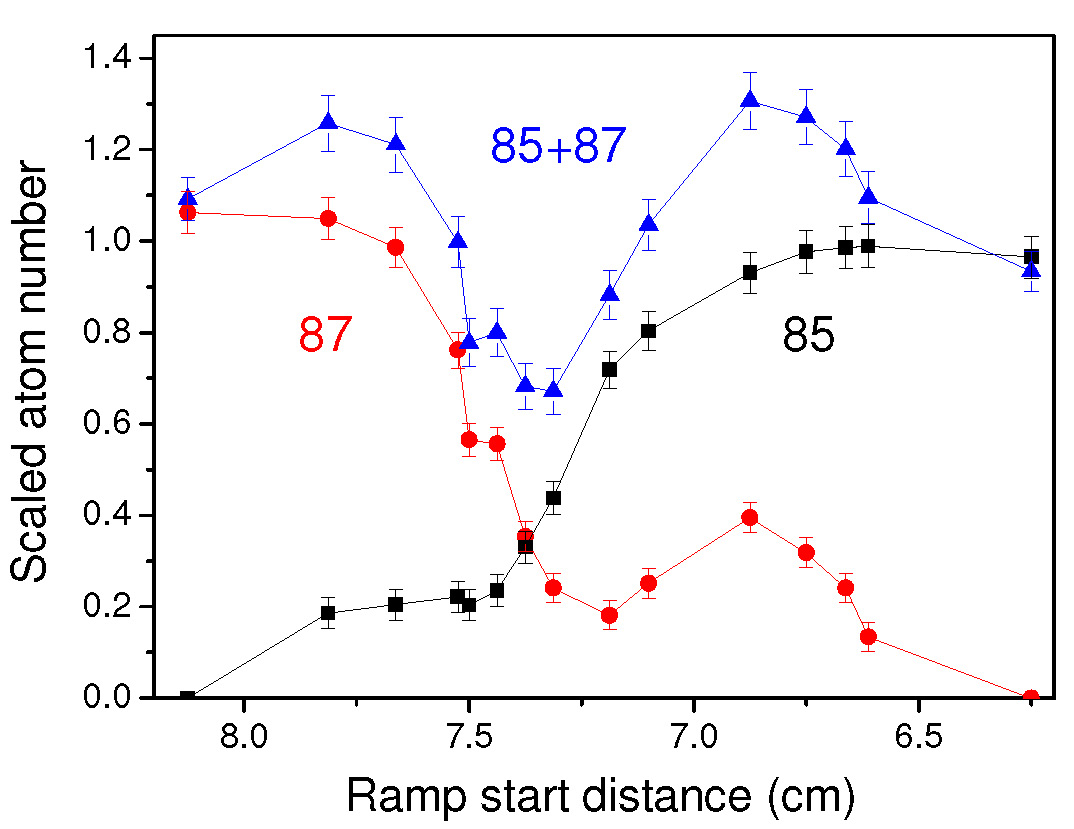}
	\caption{Merging of the two rubidium isotopes using the optimum ramp length and merging speed determined in the $^{85}$Rb experiments. Scaled atom number is shown as a function of the ramp start distance. $^{85}$Rb is held initially in trap 2 and then merged with $^{87}$Rb in trap 1.}
	\label{fig:Mix8587}
\end{figure}

In a final experiment, we demonstrated the merging of the two different rubidium isotopes as shown in figure~\ref{fig:Mix8587} by simply using the optimum ramp length determined for $^{85}$Rb. Throughout this experiment, we ensured that the number of $^{87}$Rb atoms in trap 1 prior to the merging was the same as the number of $^{85}$Rb present in trap 2. In this case we were only able to achieve an equal merge of $\approx40\,\mbox{\%}$ from each isotope, primarily due to poorer transfer of $^{87}$Rb into the combined trap. We believe this is due to a technical limitation whereby we were unable to completely compensate for the smaller magnetic moment of $^{87}$Rb ($\mu_{87}=3/4\times\mu_{85}$). To regain the same trap stiffness as for $^{85}$Rb the magnetic field gradient has to increase by 4/3, however, this was not possible due to the current limit of the power supply to the coils. We note that for sympathetic cooling it is desirable to start with a high ratio of refrigerant species to the species to be cooled e.g. $>$~30:1 ($^{87}$Rb:$^{85}$Rb)~\cite{Papp2006,Altin2010}. Hence, despite this poorer merging efficiency, we are able to create suitable conditions for sympathetic cooling of $^{85}$Rb with $^{87}$Rb, either by choosing the appropriate ramp start distance or by simply loading less $^{85}$Rb into trap 2. Specifically we have loaded $1.5\times 10^8$ $^{87}$Rb atoms with $5\times 10^7$ $^{85}$Rb atoms, enabling straightforward access to similar starting conditions to experiments that employ separate laser cooling setups for each isotope~\cite{Papp2006,Altin2010}.

\section{Discussion and conclusion}

Before concluding we mention a few points of interest regarding our experimental results. Firstly, all the experiments described have been repeated at least twice over several months and the detailed structure seen in the merging plots of figures~\ref{fig:OverviewRamps}(a),~\ref{fig:AtomnumbervsDistance}(a) and~\ref{fig:Mix8587} is reproducible. We believe this is related to the dynamics of the atoms as they cross the magnetic potential barriers outlined in our analysis in section~\ref{sec:theoreticalresults}. This demonstrates the need for precise control and synchronisation of the transport mechanism with the gradient ramps in order to achieve reproducible merging. Secondly, despite our best effort to obtain merging of $\approx100\,\mbox{\%}$ of the atoms from each trap, we were only able to achieve $\approx75\,\mbox{\%}$ of each $^{85}$Rb cloud. A possible explanation is that more energetic atoms in the cloud are colliding with the surface of our vacuum chamber. However, we have repeated the merging with initial cloud temperatures of $\approx260\,\mbox{$\mu$K}$ and $\approx70\,\mbox{$\mu$K}$ and saw no improvement in the merging percentage for the colder atoms. An alternative explanation is that we experience enhanced Majorana losses as the intermediate trap barriers are reduced to zero during the merging. The extent of the region of magnetic field where Majorana losses will occur again depends sensitively on the exact trajectory through the potential map and this fact may also contribute to the detailed structure seen in the merging plots.

In conclusion, we have demonstrated the magnetic merging of two ultracold atomic gases by the controlled overlap of two initially well-separated magnetic quadrupole traps. Theoretically we have shown that a simple 1D analysis of the combined magnetic field potential as the two traps are brought together is sufficient to identify and understand the region where merging occurs, and leads to a clear prediction of the optimum trajectory for the merging. We have verified this prediction experimentally using $^{85}$Rb and found that the final atom number in the merged trap is maximised with minimal heating by following the predicted optimum trajectory. We believe that optimal merging trajectories could be determined for any specific coil geometries and sizes by following a similar methodology. Finally we have used the magnetic merging to create controlled variable ratio atomic mixtures of the two isotopes of rubidium into a single quadrupole trap with a simple laser system for laser cooling each isotope sequentially. This offers a simple and cost effective way to achieve suitable starting conditions for sympathetic cooling of $^{85}$Rb by $^{87}$Rb towards quantum degeneracy.

\section*{Acknowledgments}
We acknowledge support from the UK Engineering and Physical Sciences Research Council (EPSRC grant EP/F002068/1) and the European Science Foundation within the EUROCORES Programme EuroQUASAR (EPSRC grant EP/G026602/1). SLC acknowledges the support of the Royal Society.

\newpage
\section*{References}
\bibliography{iopart-num}

\providecommand{\newblock}{}
\begin{thebibliography}{10}
\expandafter\ifx\csname url\endcsname\relax
  \def\url#1{{\tt #1}}\fi
\expandafter\ifx\csname urlprefix\endcsname\relax\def\urlprefix{URL }\fi
\providecommand{\eprint}[2][]{\url{#2}}

\bibitem{Anderson1995}
{Anderson} M~H, {Ensher} J~R, {Matthews} M~R, {Wieman} C~E and {Cornell} E~A
  1995 {\em Science\/} {\bf 269} 198

\bibitem{Davis1995}
{Davis} K~B, {Mewes} M, {Andrews} M~R, {van Druten} N~J, {Durfee} D~S, {Kurn}
  D~M and {Ketterle} W 1995 {\em Phys. Rev. Lett.\/} {\bf 75} 3969

\bibitem{Bradley1995}
{Bradley} C~C, {Sackett} C~A, {Tollett} J~J and {Hulet} R~G 1995 {\em Phys.
  Rev. Lett.\/} {\bf 75} 1687

\bibitem{Chin2010}
Chin C, Grimm R, Julienne P and Tiesinga E 2010 {\em Rev. Mod. Phys.\/} {\bf
  82} 1225

\bibitem{Carr2009}
{Carr} L~D, {DeMille} D, {Krems} R~V and {Ye} J 2009 {\em New J. Phys.\/} {\bf
  11} 055049

\bibitem{Lahaye2009}
{Lahaye} T, {Menotti} C, {Santos} L, {Lewenstein} M and {Pfau} T 2009 {\em Rep.
  Prog. Phys.\/} {\bf 72} 126401

\bibitem{Bloch2008}
{Bloch} I, {Dalibard} J and {Zwerger} W 2008 {\em Rev. Mod. Phys.\/} {\bf 80}
  885

\bibitem{Stenger1998}
{Stenger} J, {Inouye} S, {Stamper-Kurn} D~M, {Miesner} H, {Chikkatur} A~P and
  {Ketterle} W 1998 {\em Nature\/} {\bf 396} 345

\bibitem{Cornell1998}
{Cornell} E~A, {Hall} D~S, {Matthews} M~R and {Wieman} C~E 1998 {\em J. Low
  Temp. Phys.\/} {\bf 113}(3) 151

\bibitem{Papp2006}
Papp S~B and Wieman C~E 2006 {\em Phys. Rev. Lett.\/} {\bf 97} 180404

\bibitem{Zhang2005}
{Zhang} J, {van Kempen} E~G~M, {Bourdel} T, {Khaykovich} L, {Cubizolles} J,
  {Chevy} F, {Teichmann} M, {Tarruell} L, {Kokkelman} S~J~J~M~F and {Salomon} C
  2005 {\em Atomic Physics 19: XIX International Conference on Atomic
  Physics\/} {\bf 770} 228

\bibitem{Jing2008}
{Jing} H, {Cheng} J and {Meystre} P 2008 {\em Phys. Rev. A\/} {\bf 77} 043614

\bibitem{Herbig2003}
{Herbig} J, {Kraemer} T, {Mark} M, {Weber} T, {Chin} C, {N{\"a}gerl} H and
  {Grimm} R 2003 {\em Science\/} {\bf 301} 1510

\bibitem{Xu2003}
{Xu} K, {Mukaiyama} T, {Abo-Shaeer} J~R, {Chin} J~K, {Miller} D~E and
  {Ketterle} W 2003 {\em Phys. Rev. Lett.\/} {\bf 91} 210402

\bibitem{Durr2004}
{D{\"u}rr} S, {Volz} T, {Marte} A and {Rempe} G 2004 {\em Phys. Rev. Lett.\/}
  {\bf 92} 020406

\bibitem{Jochim2003}
{Jochim} S, {Bartenstein} M, {Altmeyer} A, {Hendl} G, {Riedl} S, {Chin} C,
  {Hecker}~{Denschlag} J and {Grimm} R 2003 {\em Science\/} {\bf 302} 2101

\bibitem{Greiner2003}
{Greiner} M, {Regal} C~A and {Jin} D~S 2003 {\em Nature\/} {\bf 426} 537

\bibitem{Zwierlein2003}
{Zwierlein} M~W, {Stan} C~A, {Schunck} C~H, {Raupach} S~M, {Gupta} S,
  {Hadzibabic} Z and {Ketterle} W 2003 {\em Phys. Rev. Lett.\/} {\bf 91} 250401

\bibitem{Ospelkaus2006}
{Ospelkaus} S, {Ospelkaus} C, {Wille} O, {Succo} M, {Ernst} P, {Sengstock} K
  and {Bongs} K 2006 {\em Phys. Rev. Lett.\/} {\bf 96} 180403

\bibitem{Sage2005}
{Sage} J~M, {Sainis} S, {Bergeman} T and {Demille} D 2005 {\em Phys. Rev.
  Lett.\/} {\bf 94} 203001

\bibitem{Danzl2008}
{Danzl} J~G, {Haller} E, {Gustavsson} M, {Mark} M~J, {Hart} R, {Bouloufa} N,
  {Dulieu} O, {Ritsch} H and {N{\"a}gerl} H 2008 {\em Science\/} {\bf 321} 1062

\bibitem{Lang2008}
{Lang} F, {Winkler} K, {Strauss} C, {Grimm} R and {Denschlag} J~H 2008 {\em
  Phys. Rev. Lett.\/} {\bf 101} 133005

\bibitem{Ni2008}
{Ni} K, {Ospelkaus} S, {de Miranda} M~H~G, {Pe'er} A, {Neyenhuis} B, {Zirbel}
  J~J, {Kotochigova} S, {Julienne} P~S, {Jin} D~S and {Ye} J 2008 {\em
  Science\/} {\bf 322} 231

\bibitem{Viteau2008}
{Viteau} M, {Chotia} A, {Allegrini} M, {Bouloufa} N, {Dulieu} O, {Comparat} D
  and {Pillet} P 2008 {\em Science\/} {\bf 321} 232

\bibitem{Deiglmayr2008}
{Deiglmayr} J, {Grochola} A, {Repp} M, {M{\"o}rtlbauer} K, {Gl{\"u}ck} C,
  {Lange} J, {Dulieu} O, {Wester} R and {Weidem{\"u}ller} M 2008 {\em Phys.
  Rev. Lett.\/} {\bf 101} 133004

\bibitem{Papp2008}
{Papp} S~B, {Pino} J~M and {Wieman} C~E 2008 {\em Phys. Rev. Lett.\/} {\bf 101}
  040402

\bibitem{Altin2010}
Altin P~A, Robins N~P, D{\"o}ring D, Debs J~E, Poldy R, Figl C and Close J~D
  2010 {\em Rev. Sci. Instrum.\/} {\bf 81} 063103

\bibitem{Thalhammer2008}
Thalhammer G, Barontini G, De~Sarlo L, Catani J, Minardi F and Inguscio M 2008
  {\em Phys. Rev. Lett.\/} {\bf 100} 210402

\bibitem{Ketterle2008}
{Ketterle} W and {Zwierlein} M~W 2008 {\em Riv. Nuovo Cimento\/} {\bf 31} 247

\bibitem{Burke1998}
{Burke Jr} J~P, {Bohn} J~L, {Esry} B~D and {Greene} C~H 1998 {\em Phys. Rev.
  Lett.\/} {\bf 80} 2097

\bibitem{Bloch2001}
Bloch I, Greiner M, Mandel O, H{\"a}nsch T~W and Esslinger T 2001 {\em Phys.
  Rev. A\/} {\bf 64} 021402

\bibitem{Cornish2000}
Cornish S~L, Claussen N~R, Roberts J~L, Cornell E~A and Wieman C~E 2000 {\em
  Phys. Rev. Lett.\/} {\bf 85} 1795

\bibitem{Roberts2001a}
Roberts J~L, Claussen N~R, Cornish S~L, Donley E~A, Cornell E~A and Wieman C~E
  2001 {\em Phys. Rev. Lett.\/} {\bf 86} 4211

\bibitem{Donley2001}
{Donley} E~A, {Claussen} N~R, {Cornish} S~L, {Roberts} J~L, {Cornell} E~A and
  {Wieman} C~E 2001 {\em Nature\/} {\bf 412} 295

\bibitem{Cornish2006}
Cornish S~L, Thompson S~T and Wieman C~E 2006 {\em Phys. Rev. Lett.\/} {\bf 96}
  170401

\bibitem{Jesper2007}
Bertelsen J~F, Andersen H~K, Mai S and Budde M 2006 {\em Phys. Rev. A\/} {\bf
  75} 013404

\bibitem{Lewandowski2003}
Lewandowski H~J, Harber D~M, Whitaker D~L and Cornell E~A 2003 {\em J. Low
  Temp. Phys.\/} {\bf 132}(5) 309

\bibitem{Migdall1985}
{Migdall} A~L, {Phillips} W~D, {Prodan} J~V, {Bergeman} T~H and {Metcalf} H~J
  1985 {\em Phys. Rev. Lett.\/} {\bf 54} 2596

\bibitem{MCCarron2008}
McCarron D~J, King S~A and Cornish S~L 2008 {\em Meas. Sci. Technol.\/} {\bf
  19} 105601

\bibitem{Gehrtz1985}
Bjorklund G~C 1980 {\em Opt. Lett.\/} {\bf 5} 15

\bibitem{Raab1987}
Raab E~L, Prentiss M, Cable A, Chu S and Pritchard D~E 1987 {\em Phys. Rev.
  Lett.\/} {\bf 59} 2631

\bibitem{Chu1985}
{Chu} S, {Hollberg} L, {Bjorkholm} J~E, {Cable} A and {Ashkin} A 1985 {\em
  Phys. Rev. Lett.\/} {\bf 55} 48

\end{thebibliography}

\end{document}